\documentclass[sigconf, nonacm]{acmart}
\usepackage{ulem}
\usepackage{enumitem}
\usepackage{algorithm}
\usepackage{algpseudocode}
\usepackage{booktabs}
\usepackage{fancyhdr}
\usepackage{pifont}
\usepackage{tikz}
\AtBeginDocument{%
  }

\setcopyright{none} 
\copyrightyear{}
\acmYear{}
\acmDOI{}

\acmISBN{}
\renewcommand\footnotetextcopyrightpermission[1]{} 
\begin{document}

\title{FT-Transformer: Resilient and Reliable Transformer with End-to-End Fault Tolerant Attention}

\author{Huangliang Dai}
\affiliation{
  \institution{University of California, Riverside}
  \city{Riverside}
  \country{USA}
}
\email{hdai022@ucr.edu}

\author{Shixun Wu}
\affiliation{
  \institution{University of California, Riverside}
  \city{Riverside}
  \country{USA}
}
\email{swu264@ucr.edu}

\author{Jiajun Huang}
\affiliation{
  \institution{University of California, Riverside}
  \city{Riverside}
  \country{USA}
}
\email{jhuan380@ucr.edu}

\author{Zizhe Jian}
\affiliation{
  \institution{University of California, Riverside}
  \city{Riverside}
  \country{USA}
}
\email{zjian106@ucr.edu}

\author{Yue Zhu}
\affiliation{
  \institution{University of California, Riverside}
  \city{Riverside}
  \country{USA}
}
\email{yzhu303@ucr.edu}

\author{Haiyang Hu}
\affiliation{
  \institution{University of California, Riverside}
  \city{Riverside}
  \country{USA}
}
\email{hhu064@ucr.edu}

\author{Zizhong Chen}
\affiliation{
  \institution{University of California, Riverside}
  \city{Riverside}
  \country{USA}
}
\email{chen@cs.ucr.edu}




\begin{abstract}
Transformer models rely on High-Performance Computing (HPC) resources for inference, where soft errors are inevitable in large-scale systems, making the reliability of the model particularly critical. Existing fault tolerance frameworks for Transformers are designed at the operation level without architectural optimization, leading to significant computational and memory overhead, which in turn reduces protection efficiency and limits scalability to larger models. In this paper, we implement module-level protection for Transformers by treating the operations within the attention module as a single kernel and applying end-to-end fault tolerance \tikz[baseline=(char.base)]{
  \node[shape=circle,draw,fill=black,text=white,inner sep=1pt] (char) {\scriptsize \textbf{1}};
}. This method provides unified protection across multi-step computations, while achieving comprehensive coverage of potential errors in the nonlinear computations \tikz[baseline=(char.base)]{
  \node[shape=circle,draw,fill=black,text=white,inner sep=1pt] (char) {\scriptsize \textbf{2}};
}. For linear modules, we design a strided algorithm-based fault tolerance (ABFT) that avoids inter-thread communication \tikz[baseline=(char.base)]{
  \node[shape=circle,draw,fill=black,text=white,inner sep=1pt] (char) {\scriptsize \textbf{3}};
}. Experimental results show that our end-to-end fault tolerance achieves up to 7.56$\times$ speedup over traditional methods with an average fault tolerance overhead of 13.9\%.

\end{abstract}

\keywords{Fault Tolerance, Transformer, Attention Mechanism, Reliability, ABFT, Deep Learning}



\maketitle

\section{Introduction}
Transformer models \cite{attention_is_all_you_need} have revolutionized the field of deep learning, demonstrating remarkable capabilities in natural language processing \cite{ref:transformer_nlp}, computer vision \cite{ref:transformer_vision}, and multimodal learning \cite{ref:transformer_multimodal}. In recent years, the rapid advancement of large language models (LLMs) has spurred extensive research on Transformers, leading to diverse specialized models, such as GPT \cite{gpt2}, BERT \cite{ref:bert} and T5 \cite{ref:roberta}. These models utilize the attention mechanism to capture long-range dependencies and complex relationships, achieving outstanding performance across a wide range of applications.

With the ongoing advancements in model capabilities, Transformers are increasingly integrated in autonomous agents \cite{ref:ai_agent} and complex reasoning \cite{ref:compex_reasoning} scenarios that involve both human interaction and real-world engagement, placing more stringent demands on inference reliability. However, Transformer inference is frequently affected by hardware errors, making it challenging to ensure computational reliability. Previous studies have shown that due to the increase in circuit density and the miniaturization of individual transistors \cite{ref:soft_error_degradation}, soft errors are inevitable in computing systems. Moreover, Transformer models often require long-duration high-load computations. For example, generating a single token in the GPT-4 model requires 560 GFLOPs and billions of tokens are produced each day \cite{ref:gpt4}. This would accelerate hardware aging and thermal cycling \cite{ref:soft_error_gpu, ref:soft_error_degradation}, leading to an increased soft error rate.

\begin{figure}[t]
\centerline{\includegraphics[scale=0.31]{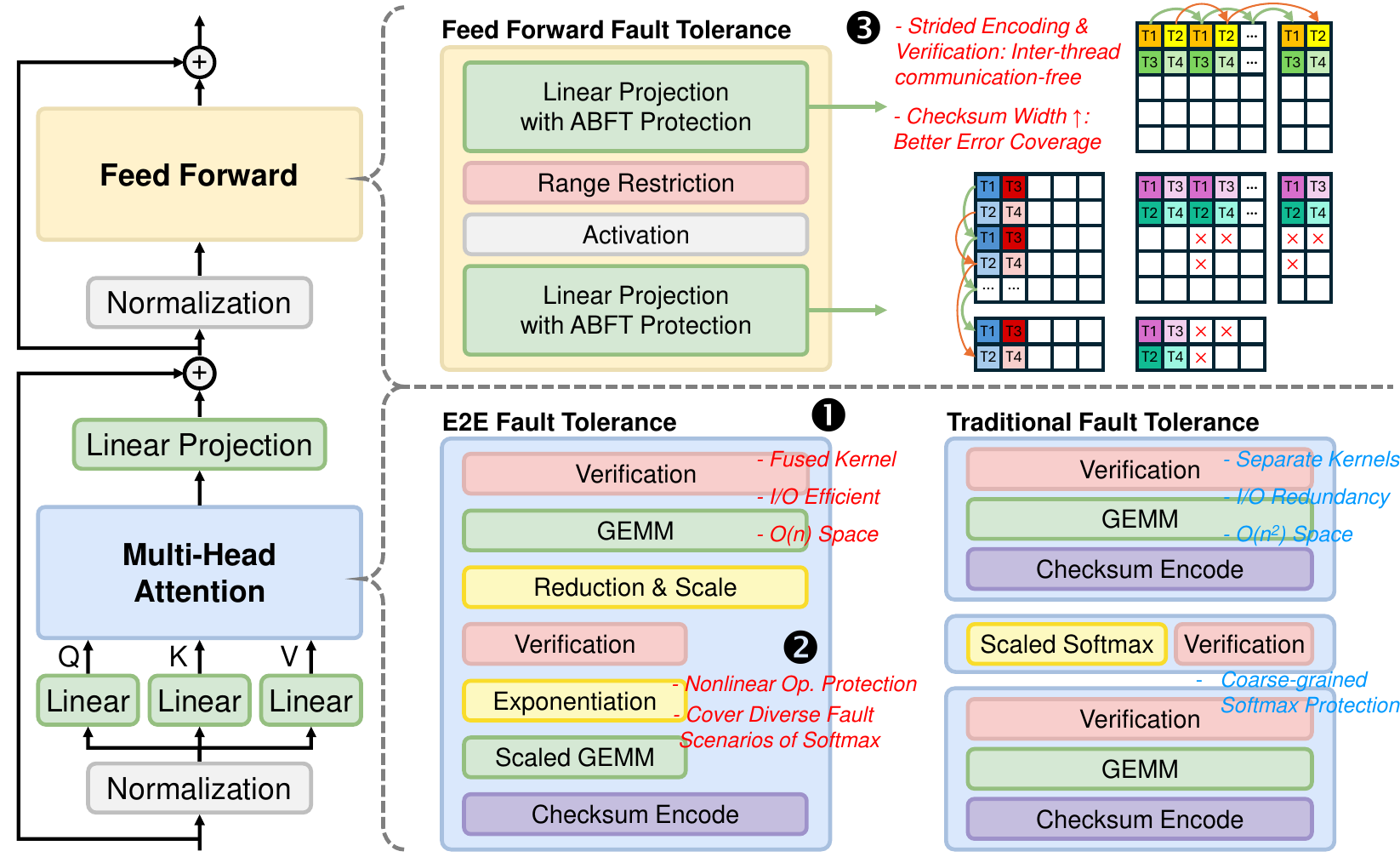}}
\caption{FT-Transformer framework, where end-to-end fault tolerant attention protects both linear and nonlinear computations within a single kernel, and the feed forward module is safeguarded using strided ABFT.}
\vspace{-5pt}
\Description{Architecture of the ft-transformer.}
\vspace{-5pt}
\label{fig:ft-transformer}
\end{figure}

Soft errors can severely affect Transformer inference, potentially leading to system crashes or silent corruption of computation results. A series of research works have investigated the impact of soft errors in accelerator hardware on deep learning model inference performance, revealing that even a single soft error can degrade prediction accuracy, with its severity influenced by data type and model architecture \cite{ref:soft_error_in_DNN, ref:dnn_resilience}. More recent studies \cite{ref:error_analysis_transformer, ref:soft_error_bert} have further explored the error resilience of Transformer models. Specifically, \cite{ref:soft_error_bert} finds that soft errors in half-precision operations used in Transformer models rarely cause overflow, but can still lead to significant degradation in inference accuracy. The operation-level error analysis in \cite{ref:error_analysis_transformer} reveals that the attention mechanism and nonlinear operations are particularly sensitive to soft errors and therefore require prioritized protection.

To maintain the reliability of Transformer computations, various fault tolerance frameworks have been introduced. In \cite{ref:error_analysis_transformer}, Xue et al. proposed a framework that integrates block-wise ABFT and range restriction to protect general matrix multiplication (GEMM) and softmax operations in vision Transformers. In \cite{ref:attn_checker}, Liang et al. designed an attention checker to detect and correct INF, NaN, and near-INF values in the GEMM operations of the attention mechanism. However, these frameworks are designed to protect individual operations, and directly applying them to Transformers incurs significant overhead, particularly for the attention mechanism. These frameworks implement fault tolerance on decoupled GEMM and softmax kernels, invoking three separate kernels to complete a single attention computation. Due to the quadratic time and memory complexity of the attention mechanism, storing intermediate computation results of GEMM and softmax would result in substantial memory consumption. Additionally, redundant memory access and excessive kernel launch overhead make these fault tolerance frameworks highly inefficient.

Driven by this motivation, we propose the FT-Transformer framework in Figure~\ref{fig:ft-transformer}. This framework adopts an end-to-end fault tolerant attention (EFTA) design \tikz[baseline=(char.base)]{
  \node[shape=circle,draw,fill=black,text=white,inner sep=1pt] (char) {\scriptsize \textbf{1}};
} that treats attention computation as a unified operation, implementing error detection and correction for GEMMs and softmax within a single kernel. The EFTA prevents memory access to $O(n^2)$ intermediate results by tiling the attention scores and performing in-place computations. This enables efficient fault tolerance for long-sequence inference. EFTA also simplifies the fault tolerance process within the fused kernel. By reordering the execution sequence of computation and fault tolerance, the three verification stages are reduced to two. Meanwhile, to enhance the protection of softmax computation, we design selective neuron value restriction (SNVR) \tikz[baseline=(char.base)]{
  \node[shape=circle,draw,fill=black,text=white,inner sep=1pt] (char) {\scriptsize \textbf{2}};
}. This method applies varying fault tolerance constraints based on computational importance, employing extended checksums to protect exponential computations and range restriction to safeguard reduction operations. Compared to traditional DMR approaches, it significantly reduces protection overhead, while offering finer fault tolerance granularity than range restriction. For linear modules, such as the feed-forward module, we design tensor checksums aligned with the thread-level data layout of tensor cores  \tikz[baseline=(char.base)]{
  \node[shape=circle,draw,fill=black,text=white,inner sep=1pt] (char) {\scriptsize \textbf{3}};
}. By performing checksum encoding and verification via strided accumulation, we realize an ABFT scheme that requires no inter-thread communication.


In this paper, the same protection mechanism is applied to both the linear modules and the linear computations within EFTA. Therefore, the following sections primarily focus on the design of EFTA. The main contributions of this paper can be summarized as follows.


\begin{itemize}[leftmargin=10pt]
\item \textbf{The first end-to-end fault tolerance framework for attention mechanism}: We design an end-to-end fault tolerant attention to protect computations against soft errors, eliminating the $O(n^2)$ memory overhead from intermediate data access and reordering computations for unified error verification across multiple steps. Our design enables reliable long-sequence inference with minimal overhead under erroneous conditions.
\item \textbf{Hybrid scheme covering all error scenarios}: The end-to-end framework restructures the computation flow and introduces rescale and reduce max operations. We propose a hybrid fault tolerance scheme that integrates strided ABFT and selective neuron value restriction (SNVR) for this paradigm to ensure comprehensive error coverage. This scheme employs checksum reuse, enabling ABFT to protect GEMM, exponentiation, and rescale operations, while applying SNVR to enhance fault tolerance in reduce-max and row-sum operations by constraining neuron values based on computational significance.
\item \textbf{Strided ABFT optimized for Tensor Cores}: To optimize fault tolerance efficiency on Tensor Cores, we develop an architecture-ware ABFT, using tensor checksum for computation verification. The strided ABFT leverages the thread-data mapping of matrix multiply-accumulate (MMA) instructions to perform strided computations, enabling intra-thread checksum encoding and error correction. With additional checksum values, this design demonstrates better fault tolerance in the presence of multiple errors. 
\item \textbf{Efficient Fault Tolerance}: We evaluate EFTA under the attention settings of medium and large models. Experimental results show that EFTA incurs an average fault tolerance overhead of 13.9\%, while achieving a 3.69$\times$ to 7.56$\times$ speedup compared to existing decoupled methods. Simulations on Transformer models indicate that EFTA is efficient in error detection and correction, incurring overheads of 4.7\% and 9.1\% respectively.
\end{itemize}


\section{Background and Related Works}
In this section, we provide an overview of the efficient attention mechanisms adopted in state-of-the-art Transformer models. We also define the fault model used in this study and review existing techniques in fault tolerant neural network design.

\subsection{Efficient Attention Mechanism Implementation}
The attention mechanism has time and space complexity of $O(n^2)$, incurring significant computational and memory overhead during inference with long sequences. To address this bottleneck and implement fast and memory-efficient inference in transformer models, several attention algorithms have been proposed, including sparse attention \cite{reformer, sparce_attention}, linear attention \cite{rnns:linear-attn, linformer}, and flash attention \cite{flashattention, flashattention2}. Among these methods, only flash attention implements standard exact attention, whereas the others achieve attention through approximation computations. This characteristic allows flash attention to be seamlessly integrated into most transformer models without altering the layer structure. The flash attention algorithm is formalized in Equations \eqref{eq2}-\eqref{eq8}. We adopt the same notations and symbols as in Subsection A, and omit the scaling factor $k_d$ for clarity in the derivation.

In flash attention, the Score tensor $\mathbf{S}$ is partitioned along the row dimension into segments $\left[\mathbf{S}^{(1)} \, \mathbf{S}^{(2)} \, \cdots \, \mathbf{S}^{(n)}\right]$, where each $\mathbf{S}^{(i)} \in \mathbb{R}^{N \times B_c}$ for $i=1,2,...,n$, and $B_c = N/n$. The Value tensor $\mathbf{V}$ is correspondingly divided along the column dimension into into $n$ segments, where each segment $\mathbf{V}^{(i)} \in \mathbb{R}^{B_r \times d}$ for $i=1, 2, \dots, n$, and $B_r = B_c$. Through this tiling approach, the global softmax can be decomposed into several local softmax operations, enabling the iterative update of the Attention tensor $\mathbf{O}$. The initial update $\mathbf{O}^{(1)}$ can be calculated as follows:
\begin{align}
\ell^{(1)} &= \text{rowsum}(e^{\mathbf{S}^{(1)}}) \in \mathbb{R}^{N} \label{eq2}\\
\widetilde{\mathbf{P}}^{(1)} &= \text{diag}(\ell^{(1)})^{-1}e^{\mathbf{S}^{(1)}} \in \mathbb{R}^{N \times B_c} \label{eq3}\\
\mathbf{O}^{(1)} = \widetilde{\mathbf{P}}^{(1)}\mathbf{V}^{(1)} &= \text{diag}(\ell^{(1)})^{-1}e^{\mathbf{S}^{(1)}}\mathbf{V}^{(1)} \in \mathbb{R}^{N \times d} \label{eq4}
\end{align}

In the second iteration, the local rowsum of $\mathbf{S}^{(2)}$ is accumulated on the previous normalization constant $\ell^{(1)}$, yielding the total rowsum for the first two segments. Scaling $\mathbf{O}^{(1)}$ and adding the local attention tensor from segment two gives $\mathbf{O}^{(2)}$.
\begin{align}
\ell^{(2)} &= \ell^{(1)} + \text{rowsum}(e^{\mathbf{S}^{(2)}}) \in \mathbb{R}^{N} \label{eq5}\\
\widetilde{\mathbf{P}}^{(2)} &= \text{diag}(\ell^{(2)})^{-1}e^{\mathbf{S}^{(2)}} \in \mathbb{R}^{N \times B_c} \label{eq6} \\
\mathbf{O}^{(2)} = \text{diag}(\ell^{(1)} &/ \ell^{(2)})^{-1} \mathbf{O}^{(1)} + \widetilde{\mathbf{P}}^{(2)}\mathbf{V}^{(2)} \in \mathbb{R}^{N \times d} \label{eq7}
\end{align}

By iteratively applying the update process, the global rowsum is aggregated across all segments, which leads to the global normalization constant $\ell^{(n)}$. The final attention tensor is then computed using this normalization factor. A detailed derivation is presented in \eqref{eq8}. Through the expansion of the scaling factors, it is demonstrated that the result produced by flash attention is equivalent to that of the standard attention. 
\begin{align}
\mathbf{O} &= \text{diag}(\ell^{(n-1)} / \ell^{(n)})^{-1} \mathbf{O}^{(n-1)} + \widetilde{\mathbf{P}}^{(n)}\mathbf{V}^{(n)} \nonumber \\
&= \text{diag}(\ell^{(n)})^{-1} \sum_{i=1}^{n-1} (e^{\mathbf{S}^{(i)}}\mathbf{V}^{(i)}) + \text{diag}(\ell^{(n)})^{-1} e^{\mathbf{S}^{(n)}}\mathbf{V}^{(n)} \nonumber \\
&= \text{diag}(\text{rowsum} (e^{\mathbf{S}}))^{-1} e^{\mathbf{S}} \mathbf{V} \in \mathbb{R}^{N \times d} \label{eq8}
\end{align}

\subsection{Fault Model}
FT-Transformer is designed to provide resilient and reliable inference against soft errors, which silently corrupt data by bit-flips and lead to incorrect inference results without any visible failure. Soft errors may arise from various hardware components, depending on the location of the fault: 1) memory faults, which affect large storage components, including main memory and cache; 2) computing unit faults, which occur in arithmetic or control logic components such as ALUs and FPUs, leading to incorrect computation outcomes; 3) interconnect faults, which arise in buses, interconnects, or NoC components and may cause data corruption or loss during transfers between compute units and memory.

This study focuses on detecting and correcting soft errors from computing unit faults, assuming that memory faults are mitigated by Error Correction Code (ECC) \cite{ref:FTinNN_ECC} and interconnect faults are managed by FT-MPI \cite{ref:ft_mpi}. Furthermore, FT-Transformer is developed based on a single event upset (SEU) assumption \cite{ref:seu}, where at most one computational error is expected during each detection and correction cycle. Due to the low occurrence rate of multiple soft errors enabled by short fault detection intervals, this assumption is considered valid and has been adopted in previous studies \cite{ref:seu_example, ref:bg_abft_kmeans}.

\subsection{Fault Tolerance in Deep Learning}
Various fault tolerance methods have been explored in deep learning to improve model robustness and reduce the impact of computational errors on prediction accuracy. These methods can be categorized as follows: 1) Modifying the model architecture, including approaches such as adding additional fault tolerant modules \cite{ref:FFinNN_change_arch1,ref:FFinNN_change_arch2,ref:FFinNN_change_arch3} or duplicating critical layers \cite{ref:FFinNN_dup_arch1,ref:FFinNN_dup_arch2,ref:FFinNN_dup_arch3,ref:FFinNN_dup_arch4}; 2) Retraining the model parameters, introducing artificial errors into the dataset or layer outputs during training. \cite{ref:FFinNN_param_retrn1,ref:FFinNN_param_retrn2,ref:FFinNN_param_retrn3}; 3) Restricting neuron values \cite{ref:FFinNN_restr2,ref:FFinNN_restr3}. In addition to these explorations in model intrinsic resilience, many methods from the high-performance computing (HPC) domain can also be applied to the fault tolerance design of learning networks. Classical approaches include dual modular redundancy (DMR) \cite{ref:FTinNN_DMR}, error-correcting code (ECC) \cite{ref:FTinNN_ECC}, and algorithm-based fault tolerance (ABFT) \cite{ref:FTinNN_ABFT, ref:bg_abft_kmeans, ref:bg_abft_fft}. Among these three approaches, ABFT is commonly adopted in deep learning models \cite{ref:FTinNN_FtCnn,ref:FTinNN_AbftinNN} due to its low overhead and compatibility with different model architectures. The fundamental principle of ABFT is to utilize redundant information encoded in checksums to verify the correctness of matrix multiplications. The encoding process can be mathematically formulated as:
\begin{align}
    \mathbf{A} \xrightarrow{\text{encode}} &\mathbf{A}^c := 
    \left [ \begin{matrix}
        \mathbf{A} \\
        \mathbf{c}_1 \mathbf{A} \\
        \mathbf{c}_2 \mathbf{A}
    \end{matrix} \right ] \in \mathbb{R}^{(M+2) \times K}  \label{eq9} \\ 
    \mathbf{B} \xrightarrow{\text{encode}} \mathbf{B}^r := 
    &\left [ \begin{matrix}
        \mathbf{B} & \mathbf{B} \mathbf{r}_1 & \mathbf{B} \mathbf{r}_2
    \end{matrix} \right ] \in \mathbb{R}^{K \times (N+2)}  \label{eq10}
\end{align}
where $\mathbf{c}_1 = \left[ 1,1,1,...,1\right]^{\top}$ and $\mathbf{c}_2 = \left[ 1,2,3,...,M\right]^{\top}$ denote the column checksum weights, while $\mathbf{r}_1 = \left[ 1,1,1,...,1\right]$ and $\mathbf{r}_2 = [ 1,2,3,...,$ $N]$ represent the row checksum weights. The encoded matrices $\mathbf{A}^c$ and $\mathbf{B}^r$ are then multiplied to produce the verification matrix $\mathbf{C}^f$, which equals to original multiplication result $\mathbf{C}$ augmented with two checksum columns $\mathbf{C}^{c1}$, $\mathbf{C}^{c2}$, and two checksum rows $\mathbf{C}^{r1}$, $\mathbf{C}^{r2}$. To detect errors in the matrix $\mathbf{C}$, its column values are weighted and summed to produce $\mathbf{C}^{c1'}=\sum_{i=0}^{M-1}\mathbf{C}_{i,0:N-1}$ and $\mathbf{C}^{c2'}=\sum_{i=0}^{M-1} (i+1) \times \mathbf{C}_{i,0:N-1}$. Then $\mathbf{C}^{c1'}$ is compared with $\mathbf{C}^{c1}$. If they are not equal to each other at position $j$, the error can be located by row index $i=\frac{\mathbf{C}^{c2'}[j]-\mathbf{C}^{c2}[j]}{\mathbf{C}^{c1'}[j]-\mathbf{C}^{c1}[j]}$ and column index $j$, and it can be corrected by adding $\mathbf{C}^{c1'}[j]-\mathbf{C}^{c1}[j]$ to element $\mathbf{C}[i][j]$. The row checksum correction follows a similar procedure. In this paper, we adopt a hybrid approach integrating ABFT and neuron value restriction in our fault tolerance scheme and redesign the detection workflow to achieve efficient end-to-end fault tolerant attention.

\section{End-to-End Fault Tolerant Attention}
\label{sec:ETOE-FT-Attn}
In this section, we present the first end-to-end fault tolerant attention (EFTA) mechanism for transformer models, achieving efficient error detection and correction within a single kernel, while mitigating potential memory faults and supporting reliable inference on long input sequences. 


\subsection{Operation-level protection for Attention with decoupled kernels}\label{subsec:FT-Attn-Baseline}
Existing fault tolerance techniques can be adapted to the attention mechanism for basic error detection and correction. The ABFT for matrix-matrix multiplication can be utilized to protect the similarity computation of tensors $\mathbf{Q}$ and $\mathbf{K}$, as well as the weighted aggregation of tensors $\mathbf{P}$ and $\mathbf{V}$, owing to their reliance on GEMM. For nonlinear operations, DMR can be utilized to verify the integrity of tensor $\mathbf{P}$ following the row softmax operation. The exponential operation and normalized weight calculation in the softmax function can be validated using the following two equations, where $n$  denotes the iteration index of the repeated computation.
\begin{align}
&\text{DMR}(e^{\mathbf{S}}) = \frac{e^{\mathbf{S}}_n + e^{\mathbf{S}}_{n-1}}{2}, \quad |e^{\mathbf{S}}_n - e^{\mathbf{S}}_{n-1}| < \epsilon \label{eq11}\\
&\text{DMR}(\mathbf{P}) = \mathbf{P}_n, \quad \, |\text{rowsum}(\mathbf{P}_n) - {\mathbf{c}_1}| < \epsilon \label{eq12} 
\end{align}

The DMR validation kernel iteratively executes these two operations until the difference between consecutive computations satisfies a predefined error tolerance. These methods safeguard computations within the attention mechanism independently. The following subsections analyze the decoupled protection framework and the corresponding fault tolerance workflow in detail.

\begin{figure}[t]
\centerline{\includegraphics[scale=0.32]{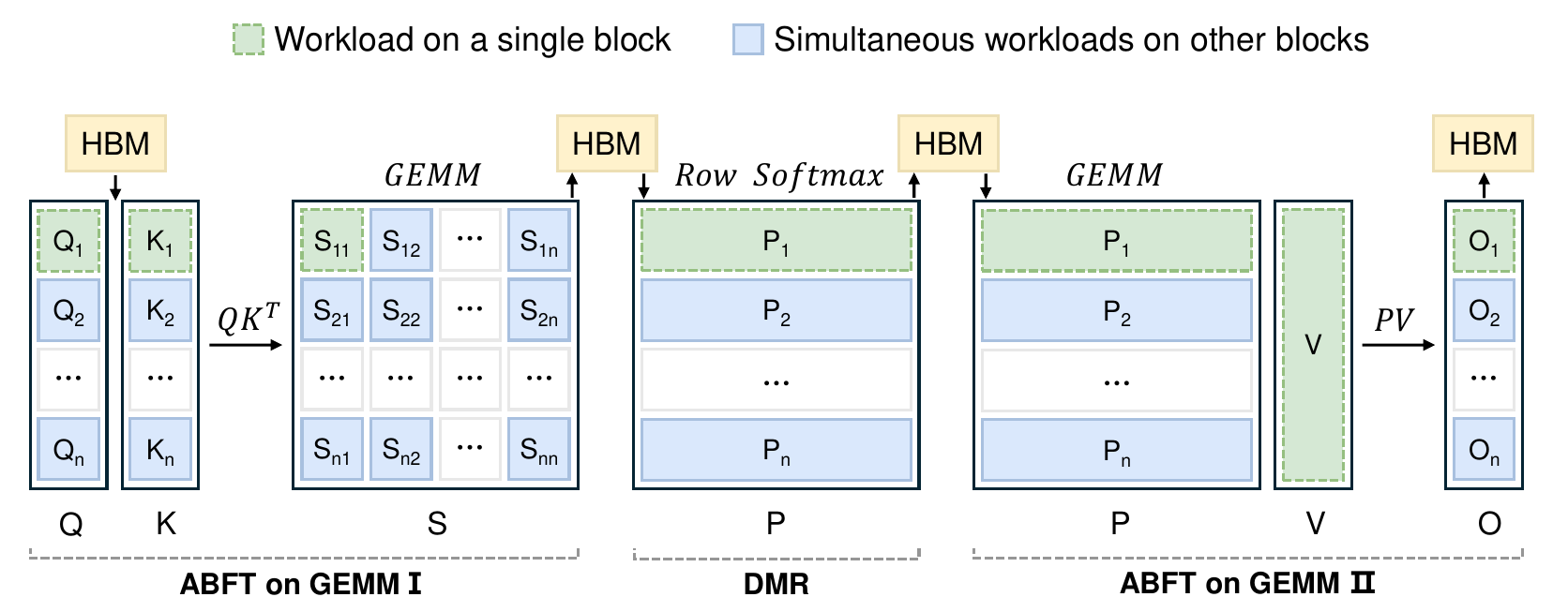}}
\caption{Decoupled fault tolerant kernels at the block level.}
\vspace{-3pt}
\Description{Decoupled fault tolerant attention}
\label{fig:ft-attn arch_base}
\end{figure}

\textbf{\large \textbullet{ }}\textbf{operation-level protection Framework.} The block-level representation of the decoupled protection framework for the attention mechanism is illustrated in Figure~\ref{fig:ft-attn arch_base}. The attention computation is divided into three kernels executed sequentially, with each kernel accessing the High Bandwidth Memory (HBM) for data reads and writes. These kernels break down computation and protection into tiled blocks, enabling finer-grained fault tolerance and in-kernel parallelism. The green blocks represent operations executed by a single cooperative thread array (CTA) on the GPU, whereas the blue blocks indicate parallel workloads distributed across other CTAs. 


All tensors involved in the attention computation have the shape $batch \times num\_head \times seq\_len \times feature\_dim$. Since the $batch$ and $num\_head$ dimensions are inherently parallel, block tiling is applied exclusively to the $seq\_len$ and $feature\_dim$ dimensions. For the ABFT on GEMM \MakeUppercase{\romannumeral 1}, $\mathbf{S}$ is divided into submatrices $\mathbf{S}_{ij}$ of size $B_r \times B_c$. Similarly, $\mathbf{Q}$, $\mathbf{K}^\top$ are tiled along the rows and columns, respectively, using the corresponding block sizes. Each time $\left\lceil \frac{seq\_len}{B_r} \right\rceil \times \left\lceil \frac{seq\_len}{B_c} \right\rceil$ blocks are processed in parallel, computing $\mathbf{Q}_i\mathbf{K}_j^\top$ with ABFT protection. For the DMR applied to row softmax, tiling is performed along the rows, with each block computing $\mathbf{P}_i$ using equations~\eqref{eq11}, \eqref{eq12}. The ABFT on GEMM \MakeUppercase{\romannumeral 2} computes $\mathbf{P}_i\mathbf{V}$ within each block.

\begin{figure}[t]
\centerline{\includegraphics[scale=0.342]{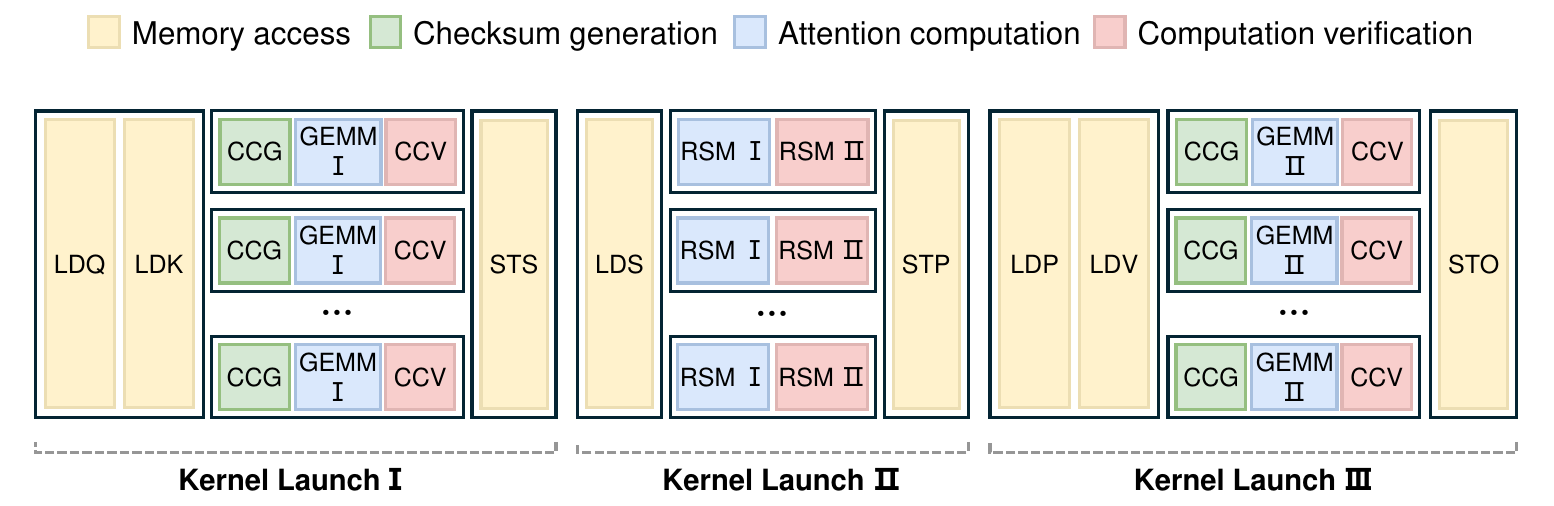}}
\caption{Workflow of the decoupled fault tolerance with operation-level protection.}
\vspace{-3pt}
\Description{Workflow of the decoupled fault tolerance}
\vspace{-4.5pt}
\label{fig:ft-attn workflow_base}
\end{figure}

\textbf{\large \textbullet{ }}\textbf{Decoupled Fault Tolerance Workflow.} Figure~\ref{fig:ft-attn workflow_base} depicts the workflow designed for fault tolerant computations within the decoupled protection framework. The yellow sections represent memory operations, with LD and ST indicating tensor data loading and storing. The green sections denote checksum encoding computations, where CCG stands for computation checksum generation. The blue sections represent original attention operations, where RSM denotes row softmax. The red sections indicate the overhead from verification, with CCV representing computation checksum verification. 


The fault tolerance workflow proceeds as follows. First, the ABFT-GEMM kernel for $\mathbf{Q}\mathbf{K}^{\top}$ is launched. Within the kernel, data blocks of $\mathbf{Q}_i$ and $\mathbf{K}_j$ are loaded from HBM to the on-chip memory of each streaming multiprocessor (SM) following the tiling scheme outlined in the decoupled protection framework. For simplicity, details of data parallelism across the $batch$ and $num\_head$ dimensions, as well as memory access patterns in shared memory and registers, are omitted. Subsequently, parallel checksum generation is executed in the tensor cores, producing row and column checksums for each block, denoted as $\mathbf{c}_1\mathbf{Q}_i$, $\mathbf{c}_2\mathbf{Q}_i$ and $\mathbf{K}_j\mathbf{r}_1$, $\mathbf{K}_j\mathbf{r}_2$. These checksums, along with the original data are then processed by the block-level GEMM operations to compute the corresponding $\mathbf{S}_{ij}$. After the computation is completed, intra-block data validation is conducted. If errors are detected, the checksums are utilized to correct computation errors. Once the correction is finalized, the result tensor $\mathbf{S}$ is written back to the HBM, and the kernel is then terminated. The DMR-RSM kernel follows a similar fault tolerance workflow, differing slightly in verification methods. It replicates the RSM module to keep successive computations within a specified tolerance range. The ABFT-GEMM kernel for $\mathbf{P}\mathbf{V}$ is a row-wise tiling variant of kernel \MakeUppercase{\romannumeral 1}.

\subsection{End-to-End Protection for Attention}
The decoupled protection framework introduces redundant memory access for tensors $\mathbf{S}$ and $\mathbf{P}$, resulting in quadratic memory footprint and increased data transfer overhead, along with the need for three kernel launches per attention computation. Inspired by flash attention, we propose an end-to-end protection framework that achieves efficient fault tolerance in a single kernel.

\begin{figure}[t]
\centerline{\includegraphics[scale=0.342]{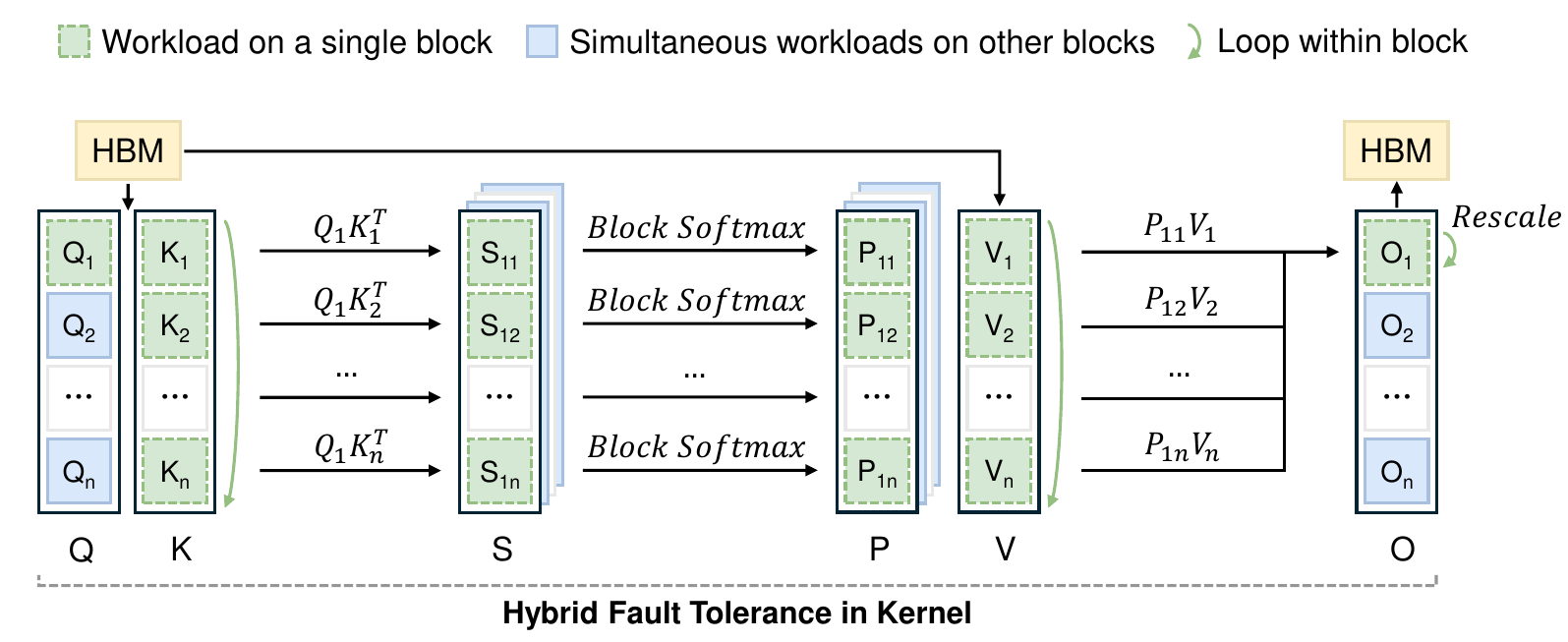}}
\caption{End-to-end fault tolerant attention at the block level.}
\vspace{-5pt}
\Description{End-to-end fault tolerant attention}
\vspace{-5pt}
\label{fig:ft-attn arch_etoe}
\end{figure}

\textbf{\large \textbullet{ }}\textbf{End-to-End Protection Framework.} As shown in Figure~\ref{fig:ft-attn arch_etoe}, the end-to-end protection framework allocates a higher computational workload to each block, increasing arithmetic intensity. Each CTA is responsible for executing all computations required to generate the corresponding $\mathbf{O}_i$. Specifically, blocked ABFT-GEMM of $\mathbf{Q}_i\mathbf{K}_j^\top$ is computed iteratively for $j$ ranging from $1$ to $n$. The resulting $\mathbf{S}_{ij}$ values are subsequently processed using block softmax to compute the local normalization weights, with selective neuron value restriction applied to enhance error resilience. Finally, $\mathbf{P}_{ij}\mathbf{V}_j$ is accumulated into $\mathbf{O}_i$, with $\mathbf{O}_i$ rescaled by the updated rowsum at each iteration. This follows the same intra-block loop order as the computation of $\mathbf{Q}_i\mathbf{K}_j^\top$, sequentially accumulating the results into $\mathbf{O}_i$. The GEMM and rescale operations in the final step are collectively safeguarded by an integrated ABFT mechanism. In the practical implementation, the stabilized softmax function is employed, which subtracts the maximum value of each row during computation to prevent numerical overflow. This requires an additional rescaling operation each time, multiplying by the ratio of the new row maximum to the previous row maximum. This additional computation is a linear scaling of the original result and can still be protected using ABFT. A detailed explanation of this procedure is provided in Section~\ref{subsec:neuron_restriction}.

\begin{figure}[t]
\centerline{\includegraphics[scale=0.42]{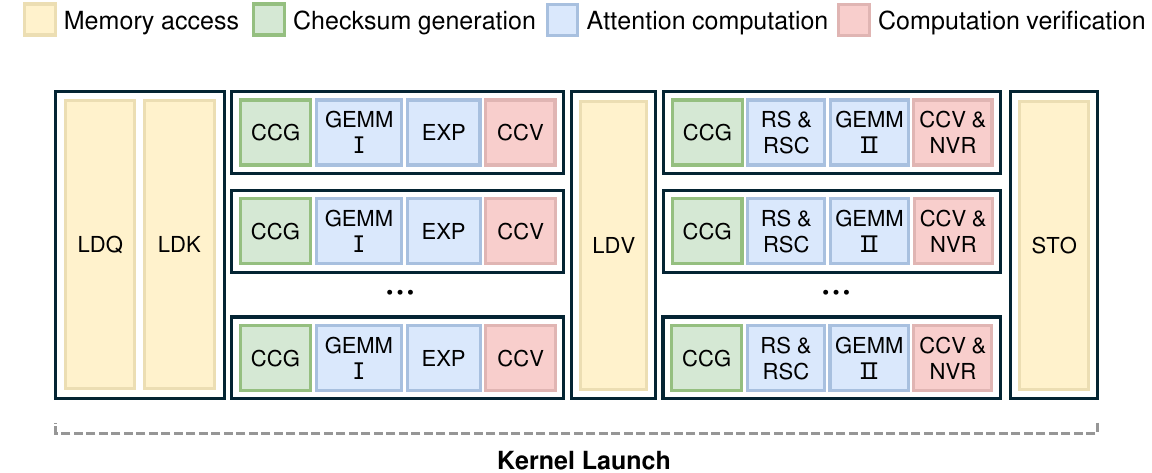}}
\caption{Workflow of the end-to-end fault tolerance with hybrid protection.}
\vspace{-5pt}
\Description{Workflow of the end-to-end fault tolerance}
\vspace{-5pt}
\label{fig:ft-attn workflow_etoe}
\end{figure}

For the block tiling, $\mathbf{Q}$, $\mathbf{K}$, and $\mathbf{V}$ are partitioned along the $seq\_len$ dimension with the same block size $B$. Since the computations of $\mathbf{Q}_i\mathbf{K}_j^\top$ and $\mathbf{P}_{ij}\mathbf{V}_j$ are executed sequentially in each block, the fused EFTA kernel sends the $seq\_len \times  num\_head \times \left\lceil \frac{seq\_len}{B} \right\rceil $ blocks to the GPU at each launch.



\textbf{\large \textbullet{ }}\textbf{End-to-End Fault Tolerance Workflow.} The workload of end-to-end fault tolerance attention is presented in Figure~\ref{fig:ft-attn workflow_etoe}. The figure illustrates the computation and fault-tolerance protection process corresponding to a single update step of $\mathbf{O}_i$, while simplifying the details of overlaps in the pipeline for clarity. The computation of block softmax is decomposed into two primary operations: the exponential function, denoted as EXP, and the row-wise summation, denoted as RS. Additionally, a rescaling operation, referred to as RSC, is applied before updating the accumulated results. The purpose of RSC is to ensure that the final output $\mathbf{O}_i$ is computed based on the global row-max and row-sum across all blocks. The NVR stands for neuron value restriction, which detects errors by ensuring computation results remain within the theoretical range of neuron outputs. This approach is widely used in deep learning \cite{ref:FFinNN_restr2,ref:FFinNN_restr3}, providing lightweight protection. 

In our end-to-end protection framework, we adopt an optimized selective NVR to impose different constraints based on the characteristics of each operation. Since the attention mechanism focuses on token positions with high similarity scores, it is crucial to maintain the relative magnitude between values. We employ ABFT to protect the GEMM of $\mathbf{Q}_i\mathbf{K}_j^\top$ and the EXP of $e^{\mathbf{S}_{ij}}$ using a unified set of checksums, ensuring precise protection for operations that determine relative numerical relationships. For the GEMM of $\mathbf{P}_{ij}\mathbf{V}_j$ and the RSC, a similar process is applied: the CCG generates input tensor checksums, which are computed alongside the original data, and the CCV verifies both operations simultaneously. For RS, which is used to compute the normalization factor and scale all values accordingly, we apply NVR with range-based protection. Moreover, by delaying detection, CCV and NVR are performed simultaneously, achieving a streamlined verification process with improved efficiency. In end-to-end attention, block softmax computes $\mathbf{P}_{ij}$ in place, reducing memory overhead and enabling long-sequence protection.

\subsection{Strided ABFT Tailored for Tensor Core}



\begin{figure}[h]
\centerline{\includegraphics[scale=0.26]{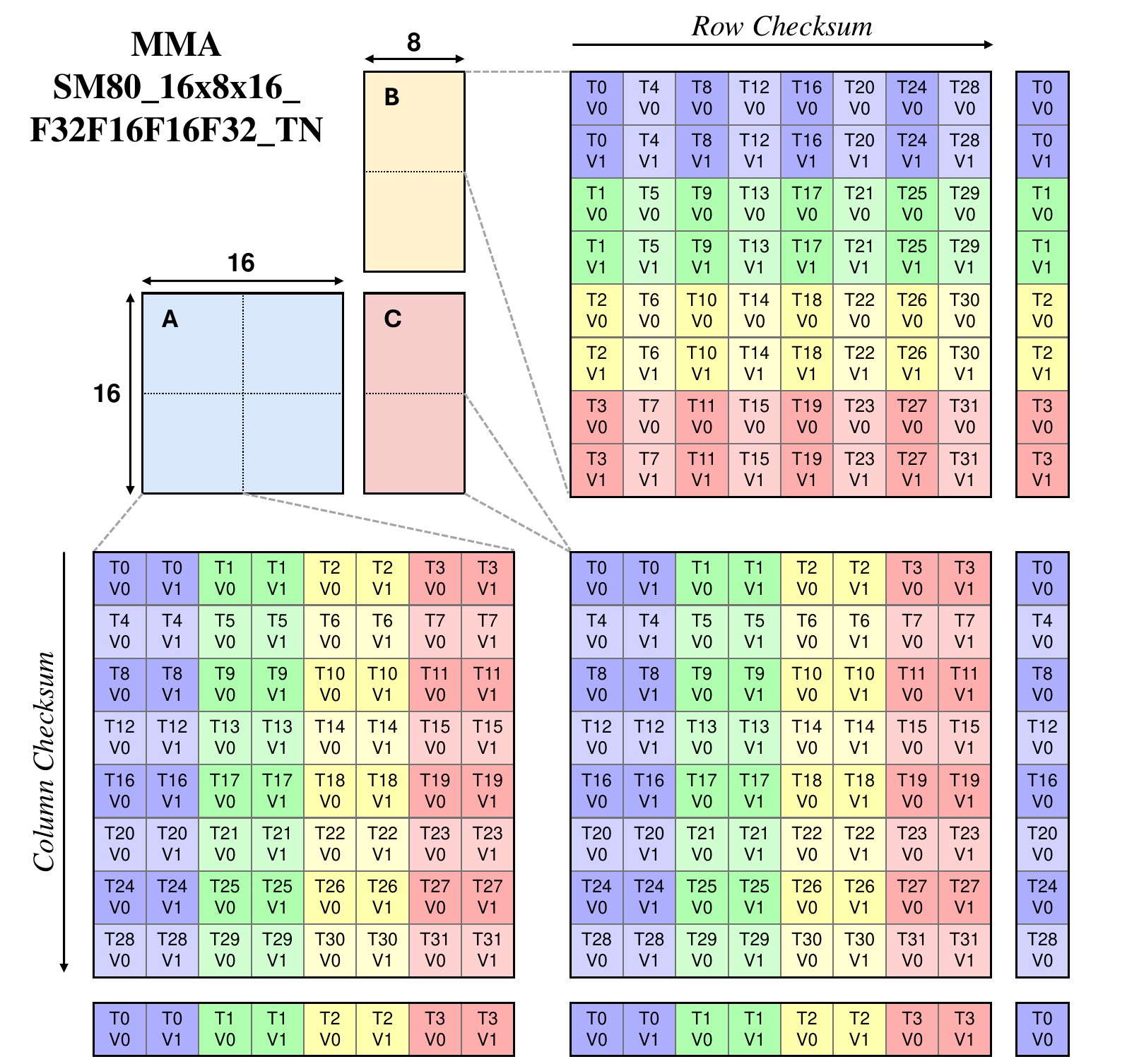}}
\caption{Thread-data layout of the SM80 16×8×16 F32F16F16F32 TN MMA instruction with conventional checksum.}
\vspace{-2pt}
\Description{Thread-data layout of the SM80\_16×8×16}
\vspace{-5pt}
\label{fig:tensor_core_arch}
\end{figure}

In our end-to-end fault tolerant attention implementation, we employ the SM80\_16×8×16\_F32F16F16F32\_TN MMA instruction for the GEMM operation, with its thread-to-data layout illustrated in Figure~\ref{fig:tensor_core_arch}. This instruction parallelizes execution across a warp of 32 threads, performing matrix multiplication on two half-precision (FP16) matrices with dimensions $M=16$, $N=8$, and $K=16$, while accumulating the results in single precision (PF32). The matrix data are distributed across all threads within the warp, with each thread contributing a portion of the registers to collectively represent the matrix. As shown in Figure~\ref{fig:tensor_core_arch}, each of the 32 threads ($T_0$ - $T_{31}$) provides two 2-byte registers $V_0$ and $V_1$, which together represent an 8×8 matrix tile. It can be observed that $\mathbf{A}[0][0]$ is stored in register $V_0$ of thread $T_0$, $\mathbf{A}[4][0]$ is stored in register $V_0$ of thread $T_{16}$, and $\mathbf{A}[8][0]$ is stored in register $V_0$ of thread $T_0$. The data corresponding to a single column is distributed across different threads, and this significantly reduces the efficiency of traditional column checksum computations of $\mathbf{A}^{c1}=\sum_{i=0}^{M-1}\mathbf{A}_{i,0:K-1}$ and $\mathbf{A}^{c2}=\sum_{i=0}^{M-1} (i+1) \mathbf{A}_{i,0:K-1}$. To compute the checksum, the data from each column must be aggregated and transferred to a single thread for accumulation, producing a $1 \times K$ tensor. This causes thread divergence and communication overhead, limiting the efficient utilization of the MMA instruction's minimal computational unit. The same issue also occurs during the computation of the row checksum. Therefore, a tensor checksum is proposed to address these issues.

\begin{figure}[t]
\centerline{\includegraphics[scale=0.337]{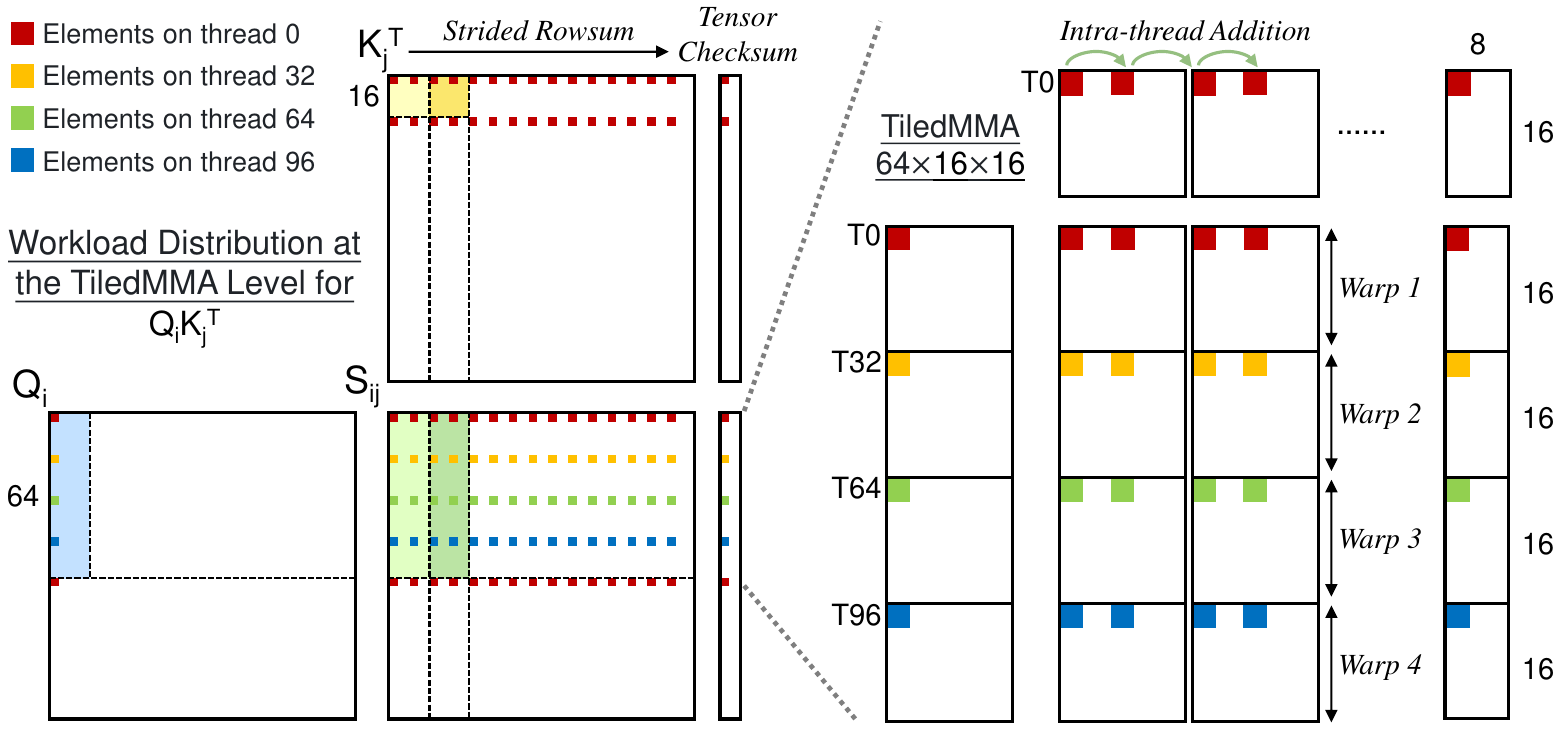}}
\caption{Workload distribution at the TiledMMA level for $Q_iK_j^T$, and the corresponding tensor checksum design.}
\vspace{-6pt}
\label{fig:tensor_checksum}
\vspace{-6pt}
\Description{Tensor checksum design}
\end{figure}

\textbf{\large \textbullet{ }}\textbf{Tensor Checksum Design.} The block level tensor checksum design is demonstrated in Figure~\ref{fig:tensor_checksum}. In this case, $\mathbf{Q}_i\mathbf{K}_j^T$ is used as an example. The computation of a block GEMM is is further divided into smaller operations by TiledMMA. The TiledMMA is constructed by combining multiple MMA Atoms of PTX instructions, enabling more complex partitioning patterns. For the end-to-end fault tolerant attention, four warps (128 threads) work in parallel to compute a 64×16×16 part of $\mathbf{S}_{ij}=\mathbf{Q}_i\mathbf{K}_j^T$, and the full block computation is covered by repeating the TiledMMA operation along the $M$, $N$, and $K$ dimensions. This 64×16×16 TiledMMA configuration increases warp-level parallelism along the $M$ dimension, replicating the workload along the $N$ dimension. As shown in Figure~\ref{fig:tensor_checksum}, along the column, elements with a stride of 64 are on the same thread. For instance, $\mathbf{Q}_i[0][0]$, $\mathbf{Q}_i[64][0]$, and $\mathbf{Q}_i[128][0]$ are stored in thread 0. Along the row, elements with a stride of 8 are on the same thread. The layout shows that $\mathbf{K}^\top_j[0][0]$, $\mathbf{K}^\top_j[0][8]$, and $\mathbf{K}^\top_j[0][16]$ are stored in thread 0. Based on this observation, the tensor checksum is designed by adding the row or column elements that are stored in the same thread at a fixed stride. The row-wise tensor checksum for any block of $\mathbf{K}^\top$ can be constructed as follows:
\begin{align}
\text{tensor checksum}_1(\mathbf{K}^\top) &= \sum_{l=0}^{lc} \mathbf{K}^\top_{:,l:l+s} \circ \mathbf{r}_1 \in \mathbb{R}^{d \times s}\label{eq13} \\
\text{tensor checksum}_2(\mathbf{K}^\top) &= \sum_{l=0}^{lc} \mathbf{K}^\top_{:,l:l+s} \circ \mathbf{r}_2 \in \mathbb{R}^{d \times s}\label{eq14} 
\end{align}
where $lc$ represents the loop count of strided addition along the row, $s$ denotes the chosen stride, and $d$ is the head dimension of tensor $\mathbf{K}^\top$. The tensor checksum weights $\mathbf{r}_1$ and $\mathbf{r}_2$ have dimensions $d \times B$, where $\mathbf{r}_1$ is an all-one tensor, and $\mathbf{r}_2$ is filled with the value $l+1$. In our end-to-end fault tolerant attention, the row stride $s$ is set to 8, since TiledMMA populates the $N$ dimension by iteratively computing the 8×8 MMA Atom. Therefore, we can derive that $lc = \left\lceil {d}/{8} \right\rceil -1$. We can encode the column tensor checksum in a similar manner. Based on the analysis above, this column checksum must be computed with a stride of 64 and should have dimensions of $64 \times d$. This will incur 8 times memory overhead than the row tensor checksum. Therefore, for the ABFT scheme applied to attention, we adopt a row-checksum-only design. The same tensor checksum encoding method is also employed for fault tolerance in $\mathbf{P}_{ij}\mathbf{V}_j$ within the attention mechanism and can be extended to mixed-precision linear operations in the feed-forward layers.

\textbf{\large \textbullet{ }}\textbf{Strided ABFT with Tensor Checksum..} With the proposed tensor checksum design, we implement an efficient algorithm-based fault tolerance mechanism for linear operations on Tensor Cores. For each block computing $\mathbf{Q}\mathbf{K}^{\top}$ in the end-to-end fault tolerance attention, two row-wise tensor checksums for $\mathbf{K}^{\top}$, denoted as $\mathbf{K}^{\top}_{check1}$ and $\mathbf{K}^{\top}_{check2}$, are encoded following Equations \eqref{eq13} and \eqref{eq14}. These checksums are appended as new columns to the original tensors, and left multiplied with $\mathbf{Q}$, producing the corresponding row checksums $\mathbf{S}^{\top}_{check1}$ and $\mathbf{S}^{\top}_{check2}$:
\begin{align}
\mathbf{S}_{check1} &= \mathbf{Q}\mathbf{K}^{\top}_{check1} = \sum_{l=0}^{lc} \mathbf{Q}\mathbf{K}^\top_{:,l:l+s} \circ \mathbf{r}_1 = \sum_{l=0}^{lc} \mathbf{S}^\top_{:,l:l+s} \circ \mathbf{r}_1 \label{eq15} \\
\mathbf{S}_{check2} &= \mathbf{Q}\mathbf{K}^{\top}_{check2} = \sum_{l=0}^{lc} \mathbf{Q}\mathbf{K}^\top_{:,l:l+s} \circ \mathbf{r}_2 = \sum_{l=0}^{lc} \mathbf{S}^\top_{:,l:l+s} \circ \mathbf{r}_2 \label{eq16}
\end{align}
where the both tensor checksum have dimensions $B_r \times s$. If no errors occur in the computation, the generated tensor checksum $\mathbf{S}_{check1}$ should maintain a strided addition relationship with tensor $\mathbf{S}$, where each element $\mathbf{S}_{check1}[i][j] = \sum_{l=0}^{lc} \mathbf{S}[i][j+s \times l]$. However, when computational errors arise, this equality no longer holds. In this case, the error can be located in the row index $i$ and column index $j + s\times \frac{\mathbf{S}_{check2}[i][j] - \mathbf{S}_{sum2}[i][j]}{\mathbf{S}_{check1}[i][j] - \mathbf{S}_{sum1}[i][j]}$ in the original tensor $\mathbf{S}$, where $\mathbf{S}_{sum1} = \sum_{l=0}^{lc} \mathbf{S}^\top_{:,l:l+s} \circ \mathbf{r}_1$ and $\mathbf{S}_{sum2} = \sum_{l=0}^{lc} \mathbf{S}^\top_{:,l:l+s} \circ \mathbf{r}_2$. The error can be corrected by adding ${\mathbf{S}_{check1}[i][j] - \mathbf{S}_{sum1}[i][j]}$ to the corresponding element. Since our tensor checksum employs an interleaved encoding scheme with additional checksum positions, the Strided ABFT exhibits stronger error detection and correction capabilities compared to single-row or single-column ABFT. Specifically, for the row tensor checksum utilized in our end-to-end fault tolerant attention, errors within each row can be detected and corrected as long as their positions are not spaced at multiples of 8. Theoretically, this approach can enhance fault tolerance by up to a factor of 8 compared to traditional ABFT. 

\subsection{Selective Neuron Value Restriction on Softmax}
\label{subsec:neuron_restriction}
\begin{figure}[t]
\centerline{\includegraphics[scale=0.4]{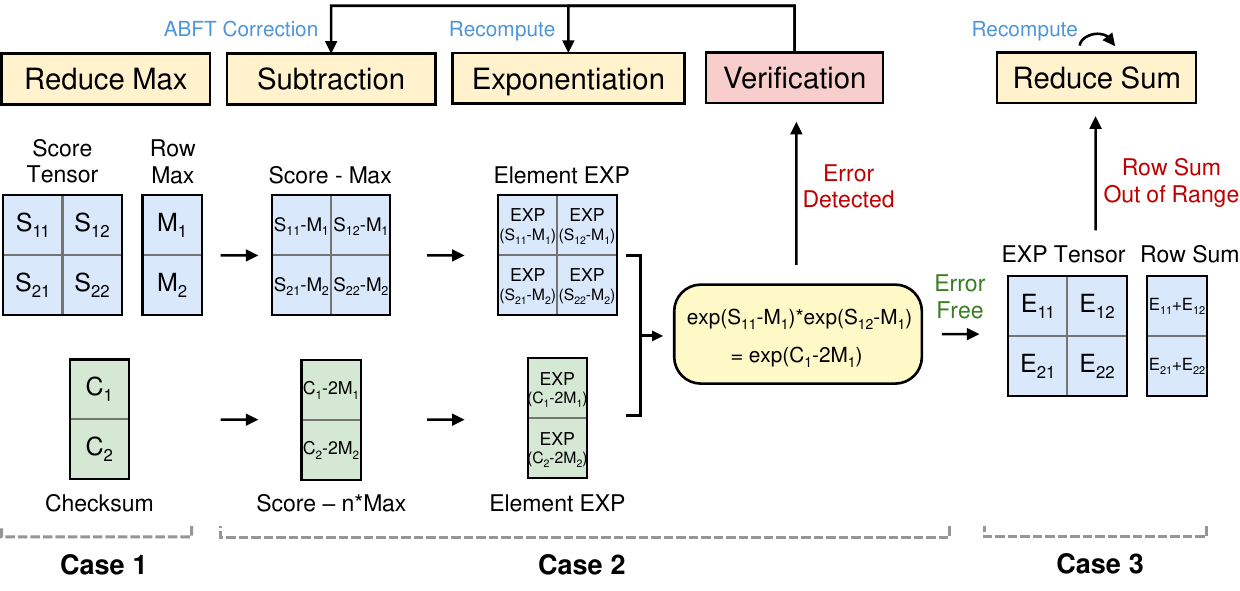}}
\caption{Selective neuron value restriction for softmax computation, covering all potential error scenarios.}
\vspace{-7pt}
\label{fig:selective_restriction}
\vspace{-8pt}
\Description{selective neuron value restriction}
\end{figure}

In softmax computations, the exponential operation determines the relative magnitudes of the output values, playing a crucial role in establishing inter-token dependency in the attention mechanism. The normalization step ensures that the results remain within a stable numerical range, which can be protected using range-based approaches. The proposed selective neuron value restriction (SNVR) method leverages the invariance of computational relationships to apply precise protection to exponential operations, while constraining normalization results within the theoretical range. This approach enables fine-grained fault tolerance in softmax computations, balancing reliability and efficiency. The SNVR is designed based on the single-event upset (SEU) assumption, where only one error is considered in each scenario. Additionally, to mitigate numerical overflow, this work employs the numerically stable softmax formulation: $p_{i,j} = \frac{\text{exp}(s_{i,j} -  s_{i\_max})}{\sum_k \text{exp}(s_{i,k}-s_{i\_max})}$, where $s_{i,j}$ denotes an element of the tensor $\mathbf{S}$, and $s_{i\_max}$ represents the maximum value in the $i\text{-}th$ row. Figure~\ref{fig:selective_restriction} illustrates the fault tolerance mechanism of selective neuron value restriction under different computational errors using a simplified checksum. Below, We detail the fault tolerance process for each case based on the actual tensor checksums.

\uline{Case 1}: An error occurs during the search for the maximum row value. If the error occurs in row $i$, the rowmax $s_{i\_max}$ is altered to $s'_{i\_max}$. During the computation of the numerator, this error propagates across the entire row, corrupting $\text{exp}(s_{i,j} - s_{i\_max}')$. Similarly, in the computation of the denominator, the error also contaminates the respective value $\sum_k \text{exp}(s_{i,k}-s_{i\_max}')$. However, such errors would not influence the final result, as the error terms $e^{s'_{i\_max}}$ are canceled out.

\uline{Case 2}: The calculation of $d_{{i,j}} = s_{i,j} - s_{i\_max}$ is affected by a computational fault or the exponential computation $\text{exp}(d_{i,j})$ is erroneous. In this case, the result of these two steps is corrupted to $(e^{{s_{i,j} - s_{i\_max}}})'$. This one-point error would further propagate to the $i\text{-}th$ row of the GEMM computation $e^{(\mathbf{S}-\mathbf{S}_{max})}V$ and the $i\text{-}th$ rowsum value $\sum_k \text{exp}(s_{i,k}-s_{i\_max})$. Therefore, this error need be corrected before these subsequent computations. The Strided ABFT can be extended to protect the max value deduction and the exponentiation by reusing the checksum when calculating the tensor $\mathbf{S}$. When we conduct element-wise deduction, the same operation is applied on the checksum, and the equality between the checksum and original data becomes $\mathbf{S}_{check1}[i][j] - (lc+1)s_{i\_max} = \sum_{l=0}^{lc} \mathbf{S}[i][j+s \times l] - s_{i\_max}$. Similarly, after the exponential computation, the checksum can be verified using $\text{exp}(\mathbf{S}_{check1}[i][j] - (lc+1)s_{i\_max}) = \prod_{l=0}^{lc} \text{exp}(\mathbf{S}[i][j+s \times l] - s_{i\_max})$. If this equation is not satisfied in the verification process, it indicates the occurrence of a computational error. For error correction, if the error arises in the linear operations of deduction, it can be directly corrected using checksum-based recovery; if the error occurs in exponentiation, we employ recomputation for correction.

\uline{Case 3}: When computing the denominator of the softmax function, errors may occur during the reduce sum operation, causing the $i\text{-}th$ rowsum to be incorrectly computed as $\sum'_k \text{exp}(s_{i,k}-s_{i\_max})$. This leads to errors in all elements of the $i\text{-}th$ row during the final computation of $\mathbf{P}$, as the normalization process scales the entire row by an incorrect factor. However, since this error does not affect the relative numerical relationships, the attention mechanism still focuses on the positions with the highest values. Therefore, it is sufficient to ensure that these values remain within the theoretical computational range, preventing significant deviations or anomalies. In practice, we constrain the rowsum within a predefined range using the following inequality: $\sum \text{exp}(s_{block\_max} - s_{global\_max}) < rowsum < n$. If the computed rowsum falls outside this range, it indicates a computational error, which is then corrected through recomputation.

\small
\begin{algorithm}[t]
\caption{EFTA with Unified Verification}
\label{alg:EFTA Algorithm}
\begin{algorithmic}[1]
\Procedure{End-to-End Fault Tolerant Attention}{}
    \State Tile $\mathbf{Q}$, $\mathbf{K}$, $\mathbf{V}$ along $seq\_len$ dimension into $blk = \left\lceil \frac{seq\_len}{B} \right\rceil$ blocks,
    \State where $\mathbf{Q}_i$, $\mathbf{K}_j$, $\mathbf{V}_j \in \mathbb{R}^{B \times d}$
    \For{$ i \;\text{from}\;1\;\text{to} \;blk $}
        \State Load $\mathbf{Q}_i$ from HBM to Tensor Core shared memory
        \For{$ j \;\text{from}\;1\;\text{to} \;blk$}
            \State Load $\mathbf{K}_j, \mathbf{V}_j$ from HBM to Tensor Core shared memory
            \State Encode tensor checksum $\mathbf{K}_j^{c_1}, \mathbf{K}^{c_2}_j, \mathbf{V}^{c_1}_j, \mathbf{V}^{c_2}_j$
            \State GEMM I: $\; \mathbf{S}_{ij} = \mathbf{Q}_i \mathbf{K}_j^T, \;\mathbf{S}^{c_1}_{ij} = \mathbf{Q}_i (\mathbf{K}^{c_1}_j)^T, \;\mathbf{S}^{c_2}_{ij} = \mathbf{Q}_i (\mathbf{K}^{c_2}_j)^T $
            \State Reduce Max: $\;m_{ij} = \max(m_{i(j-1)}, \operatorname{rowmax}(\mathbf{S}_{ij})) $
            \State Element-wise EXP: $\;\mathbf{P}_{ij} = \exp(\mathbf{S}_{ij} - m_{ij})$
            \State $\quad\quad\quad\quad\quad\quad\quad\quad\;\mathbf{P}^{c_1}_{ij}= \exp(\mathbf{S}^{c_1}_{ij}- \left\lceil \frac{d}{8} \right\rceil m_{ij})$
            \If{$|\prod \mathbf{P}_{ij} - \prod \mathbf{P}^{c_1}_{ij}| > \epsilon_1 $}
                \State Correct linear operations with tensor checksum $\mathbf{S}^{c_1}_{ij}, \mathbf{S}^{c_2}_{ij}$
                \State Correct EXP with recomputation
            \EndIf
            \State Reduce Sum: $\;\ell_{ij} = e^{m_{i(j-1)} - m_{ij}} \, \ell_{i(j-1)} + \operatorname{rowsum}(\mathbf{P}_{ij})$
            \State  GEMM II: $\;\mathbf{O}_i = \operatorname{diag}(e^{m_{i(j-1)} - m_{ij}}) \mathbf{O}_i + \mathbf{P}_{ij} \mathbf{V}_j$
            \State $\quad\quad\quad\quad\;\,\,\mathbf{O}^{c_1}_i = \operatorname{diag}(e^{m_{i(j-1)} - m_{ij}}) \mathbf{O}^{c_1}_i + \mathbf{P}_{ij} \mathbf{V}^{c_1}_j$
            \State $\quad\quad\quad\quad\;\,\,\mathbf{O}^{c_2}_i = \operatorname{diag}(e^{m_{i(j-1)} - m_{ij}}) \mathbf{O}^{c_2}_i + \mathbf{P}_{ij} \mathbf{V}^{c_2}_j$
        \EndFor
        \If{$ \ell_{ij} < \sum_k e^{(m_{ik}-m_{ij})} \; \text{or} \; \ell_{ij} > seq\_len$}
                \State $\ell_{ij} = \sum_k e^{(m_{ik}-m_{ij})}$
        \EndIf
        \State Normalization: $\;\mathbf{O}_i = \operatorname{diag}(\ell_{ij})^{-1} \mathbf{O}_i, \;\mathbf{O}^{c_1}_i = \operatorname{diag}(\ell_{ij})^{-1} \mathbf{O}^{c_1}_i$
        \State $\quad\quad\quad\quad\quad\quad\;\;\,\mathbf{O}^{c_2}_i = \operatorname{diag}(\ell_{ij})^{-1} \mathbf{O}^{c_2}_i$
        \If{$|\sum \mathbf{O}_{i} - \sum \mathbf{O}^{c_1}_{i}| > \epsilon_2 $}
                \State Correct linear operations with tensor checksum $\mathbf{O}^{c_1}_{i}, \mathbf{O}^{c_2}_{i}$
        \EndIf
    \EndFor
\EndProcedure
\end{algorithmic}
\end{algorithm}
\normalsize

\textbf{\large \textbullet{ }}\textbf{Optimized EFTA with Unified Verification.} To further reduce the overhead of fault tolerance, a checksum reuse strategy is adopted to keep track of the invariant numerical relationship across multiple computations, allowing error detection through a single verification step, as presented in Algorithm~\ref{alg:EFTA Algorithm}. The tensor checksum $\mathbf{S}^{c_1}_{ij}$ generated in GEMM I is subsequently processed through subtraction and element-wise EXP operations, integrating the information of the three computations into a unified checksum. Errors can be detected by verifying the equality of element-wise products between the original computation result $\mathbf{P}_{ij}$ and the corresponding tensor checksum $\mathbf{P}^{c_1}_{ij}$. Since the tensor $\mathbf{P}_{ij}$ is used in place during the GEMM II computation, error verification must be performed at each iteration to prevent error propagation in subsequent calculations. For GEMM II, rescale, and normalization operations, we adopt a similar unified verification strategy. Using the linearity of these three computations, we apply the same transformations to the tensor checksum $\mathbf{O}^{c_1}_{i}$, ensuring that its element-wise summation remains consistent with that of the original tensor $\mathbf{O}_{i}$. As the final result $\mathbf{O}_i$ is not fused with subsequent computations, the numerical relationship between the tensor checksum $\mathbf{O}^{c_1}_{i}$ and $\mathbf{Q}_i$ can be preserved throughout updates. This allows verification to be performed only once, after all iterations have been completed.

We also optimize the error detection process for the reduce sum operation. Rather than verifying the range of rowsum $\ell_{ij}$ after each reduce sum operation, we instead validate the final value of $\ell_{ij}$ prior to normalization, ensuring that it remains within the theoretically expected bounds. Moreover, we correct errors in rowsum by replacing them with the approximation result of the normalization factor, thereby mitigating the overhead of recomputation. In practical implementation, this approximation value is derived from the computation of $\sum_k e^{(m_{ik}-m_{ij})}$, where $m_{ik}$ is the maximum row value in each iteration and $m_{ij}$ is the global rowmax. This approximate correction still ensures reliable inference, as attention primarily focuses on the most important positions, with most values concentrated around the largest ones, while the rest have a negligible impact on the summation.



\section{Experimental Evaluation}
\label{sec:expmt}
We evaluate our end-to-end fault tolerant attention (EFTA) on a 40GB A100-PCIE GPU. The GPU device is connected to a node with one 64-core AMD EPYC 7763 CPU whose boost frequency is 3.5 GHz. We compile programs using CUDA 12.4 with optimization flag O3. The input and output tensors for EFTA are in half-precision floating-point format (FP16), and SM80 16×8×16 F32F16F16F32 TN MMA instruction is utilized for mixed-precision matrix computations on the tensor core. We first present benchmark results comparing EFTA with existing decoupled approaches in different attention configurations. Next, we assess the fault tolerance overhead and the error coverage of our hybrid scheme. Finally, we simulate the fault tolerance performance of the optimized EFTA on common Transformer models.

\subsection{End-to-End Fault Tolerance Framework for Attention Mechanism}
\label{sec:5.1}
In this subsection, we measure the runtime of EFTA and existing decoupled fault tolerance method and conduct an overhead breakdown analysis. We compare their performance across sequence lengths ranging from 512 to 16K, adjusting the batch size to maintain a total token count of 16K. Our experiments evaluate attention performance and the corresponding fault tolerance overhead under two hidden dimension settings. Based on common configurations of medium and large models, the hidden dimension of attention is set to 1024 (head = 16, dim = 64) and 4096 (head = 32, dim = 128).

\begin{figure}[t]
\centerline{\includegraphics[scale=0.58]{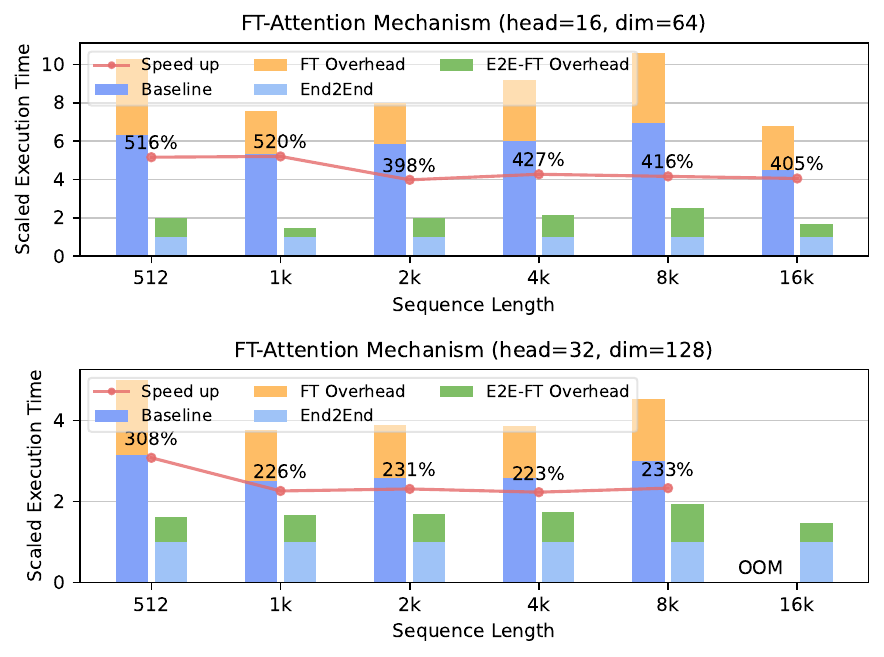}}
\vspace{-5pt}
\caption{Scaled execution time and fault tolerance overhead of end-to-end FT attention and decoupled FT attention.}
\vspace{-5pt}
\label{fig:etoe_framework_exp}
\vspace{-5pt}
\Description{etoe_framework_exp}
\end{figure}

\textbf{\large \textbullet{ }}\textbf{End-to-End Fault Tolerance Performance.} Figure~\ref{fig:etoe_framework_exp} shows the performance of the end-to-end fault tolerance framework compared to the traditional decoupled fault tolerance framework in attention computation. For attention computation in the medium model with hidden dimension = 1024, our framework achieves an average 447\% speedup over the traditional method under fault tolerance protection. Notably, for short inputs with sequence lengths of 512 and 1k, the speedup exceeds fivefold. For the large-scale model configuration with hidden dimension = 8192, our framework achieves an average 244\% speedup, and when sequence length = 512, the speedup can reach threefold. Additionally, it can be observed that the traditional method encounters an out-of-memory (OOM) error when the sequence length reaches 16k. This occurs because the traditional approach stores the intermediate results of the $\mathbf{Q}\mathbf{V}^\top$ matrix computation, requiring $batch \times num\_head \times {seq\_len}^2$ memory access to HBM. Our end-to-end method mitigates this issue by partitioning the intermediate data and incrementally accumulating the results, ensuring fault tolerance in long-sequence inference.

\begin{figure}[t]
\centerline{\includegraphics[scale=0.44]{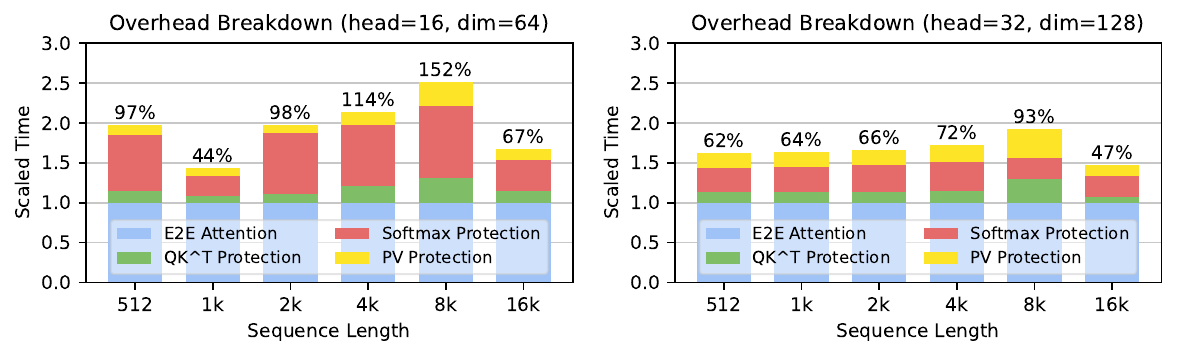}}
\vspace{-5pt}
\caption{Breakdown analysis of fault tolerance overhead in the end-to-end FT attention.}
\vspace{-5pt}
\label{fig:overhead_breakdown}
\vspace{-5pt}
\Description{overhead_breakdown}
\end{figure}

\textbf{\large \textbullet{ }}\textbf{FT Overhead Breakdown Analysis.} The detailed fault tolerance overhead breakdown analysis of EFTA is provided in Figure~\ref{fig:overhead_breakdown}.

The overhead analysis indicates that although our EFTA achieves a significant speedup over existing approaches, the overhead remains significant. For the attention mechanism of large-scale models, EFTA incurs an average fault tolerance overhead of 68\%. For medium-scale models, the average overhead is 96\%. Such high overhead stems from the incompatibility of traditional methods with EFTA's workflow and computing platform. EFTA optimizes the computation pipeline by parallelizing operations across phases, but DMR's redundant computations cannot be efficiently integrated into the fused kernel, requiring a separate phase for fault tolerance and causing 47\% average overhead for softmax protection. Meanwhile, traditional ABFT, designed for single and double precision computations, lacks efficient optimizations for mixed-precision computations on Tensor Core, leading to 35\% average overhead for GEMM protection. We develop a hybrid scheme to solve this issue, achieving efficient architecture-aware fault tolerance.



\subsection{Hybrid Scheme for Efficient Fault Tolerance}
The same experimental setup is adopted as in Section~\ref{sec:5.1}. We evaluate the fault tolerance overhead of Strided ABFT and selective neuron value restriction under different hidden dimension settings and assess their error coverage capability.

\begin{figure}[t]
\centerline{\includegraphics[scale=0.58]{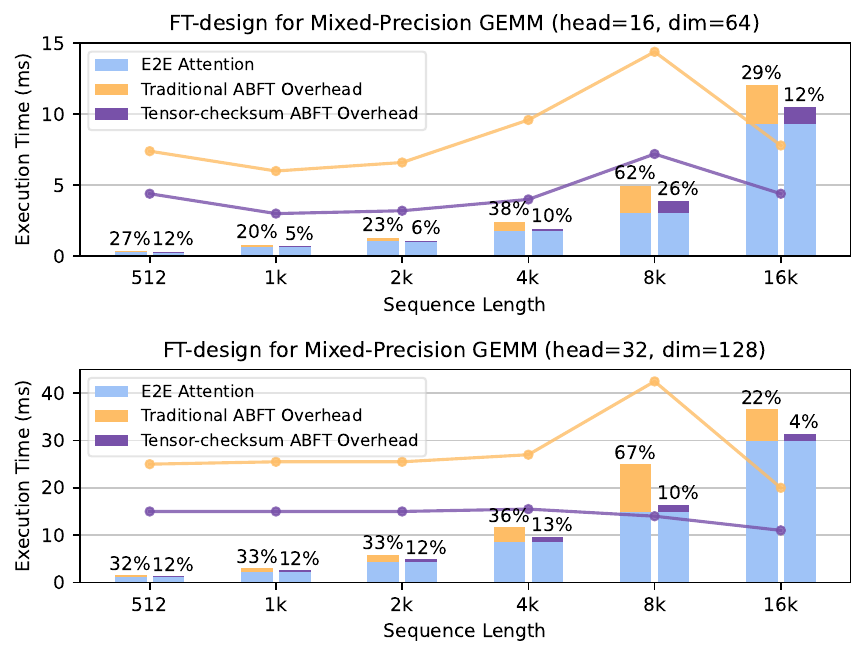}}
\vspace{-5pt}
\caption{End-to-end FT attention execution time: comparison of Strided ABFT and traditional ABFT.}
\vspace{-5pt}
\label{fig:tensor_check_final}
\vspace{-5pt}
\Description{tensor_check_final}
\end{figure}

\begin{figure}[t]
\centerline{\includegraphics[scale=0.44]{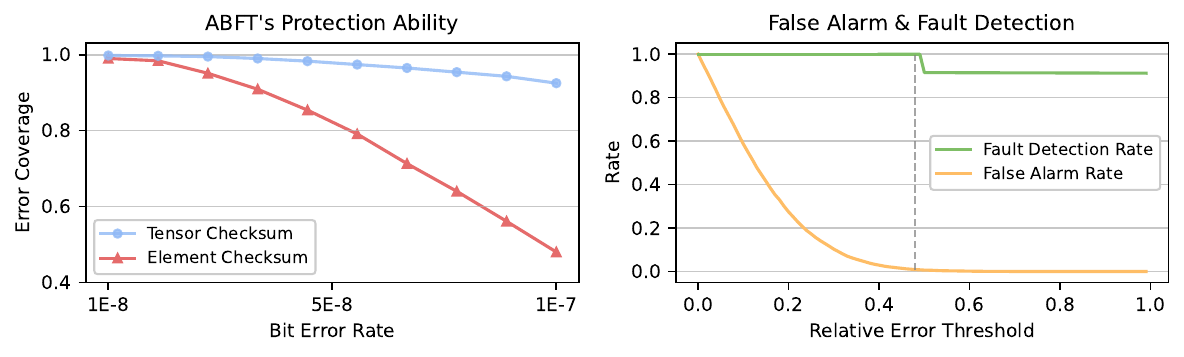}}
\vspace{-5pt}
\caption{Error coverage and false alarm rate analysis of Strided ABFT.}
\vspace{-5pt}
\label{fig:ABFT_coverage}
\vspace{-4pt}
\Description{ABFT_coverage}
\end{figure}

\textbf{\large \textbullet{ }}\textbf{EFTA with Strided ABFT.} Figure~\ref{fig:tensor_check_final} compares the performance of EFTA using Strided ABFT and traditional ABFT. The execution time for these two ABFT methods is represented by the purple and orange bars respectively, and the overhead is calculated by $\frac{fault \;tolerance \; time}{attention \; execution \;time} $ and is displayed at the top of the bars. In the experiment, the ABFT is used to protect $\mathbf{Q}\mathbf{K}^\top$ and $\mathbf{P}\mathbf{V}$ computations, and results show that with head = 16, dim = 64, and a total input token count of 16K, Strided ABFT incurs an average overhead of 11.8\% in EFTA. This reduces approximately 64\% fault tolerance overhead compared to traditional ABFT. The line plot in the figure illustrates the relative overhead of the two methods. It shows that for sequence lengths ranging from 512 to 16K, Strided ABFT consistently achieves lower fault tolerance overhead. For head = 32, dim = 128, and a total input token count of 16K, the average fault tolerance overhead of Strided ABFT is 10.5\%, which still outperforms the traditional approach across all the test sequence lengths.

\begin{figure}[t]
\centerline{\includegraphics[scale=0.58]{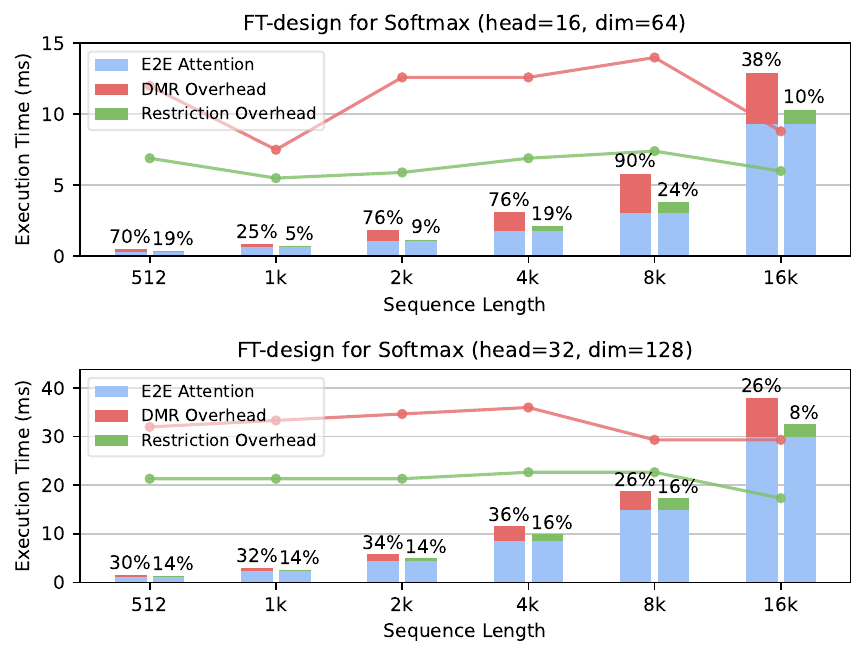}}
\vspace{-5pt}
\caption{End-to-end FT attention execution time: comparison of DMR protection and selective neuron value restriction.}
\vspace{-5pt}
\label{fig:slec_restric}
\vspace{-5pt}
\Description{slec_restric}
\end{figure}

\begin{figure}[t]
\centerline{\includegraphics[scale=0.44]{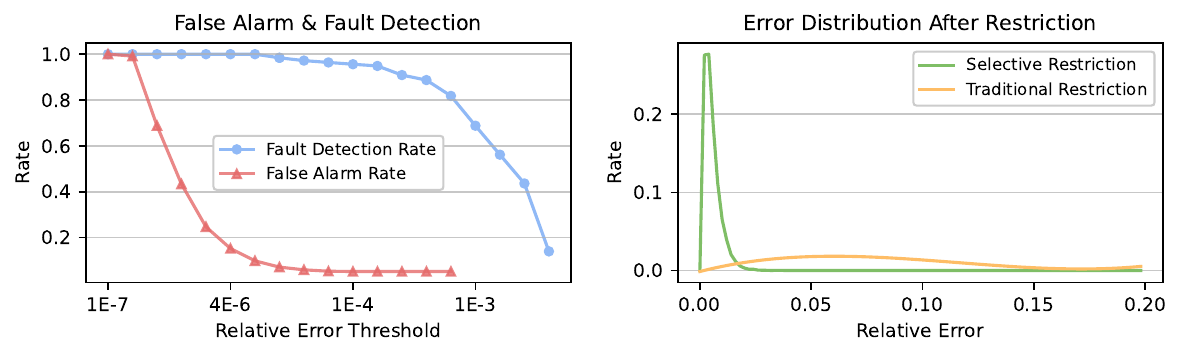}}
\vspace{-5pt}
\caption{False alarm rate and error distribution analysis of selective neuron value restriction.}
\vspace{-5pt}
\label{fig:SNVR_coverage}
\vspace{-5pt}
\Description{SNVR_coverage}
\end{figure}

\textbf{\large \textbullet{ }}\textbf{Error Coverage of Strided ABFT.} In Figure~\ref{fig:ABFT_coverage}, the left plot illustrates the error detection capability of traditional element checksum-based ABFT and our Strided ABFT. The right plot presents the true detection rate and the false detection rate of Strided ABFT under different error threshold settings. Strided ABFT is optimized according to the data-thread layout of tensor cores, employing an 8-element-wide checksum design that aligns with MMA instructions. This approach improves error coverage compared to the single-element-wide checksum used in traditional ABFT. When the computational bit error rate is 1e-7, Strided ABFT achieves an error coverage rate of 92.5\%, whereas traditional checksum achieves only 48\%. Due to intrinsic rounding errors of half-precision tensor core computations and ABFT's summation-based detection, checksum results may differ from original outputs under error-free conditions. Therefore, in practical error detection, a true error is identified when the difference between the checksums and original results exceeds the error threshold. Our experiments show that when the threshold is below 0.4, intrinsic errors are misidentified as true errors, while a threshold above 0.5 leads to some computational errors being missed. The error threshold of 0.48 gives the best detection performance.

\textbf{\large \textbullet{ }}\textbf{EFTA with Selective Neuron Value Restriction.} Within the EFTA framework, we employ DMR and selective neuron value restriction separately to protect softmax computation and compare their performance. We measure the total execution time and the fault tolerance time for both methods and derive the corresponding overhead. The experimental results are presented in Figure~\ref{fig:slec_restric}. Medium-scale models typically use attention with 1024 hidden dimensions. In this setting, selective neuron value restriction incurs an average fault tolerance overhead of 14.3\%, whereas DMR results in a significantly higher overhead of 62.5\%. For large model settings with a hidden dimension of 4096, selective neuron value selection incurs an average overhead of 13.6\%, reducing overhead by half compared to DMR's 30.6\%. The relative overhead line in the figure shows that our selective neuron value restriction achieves better fault tolerance performance for all test sequence length.

\textbf{\large \textbullet{ }}\textbf{Error Coverage of Selective Neuron Value Restriction.} We further explore the SNVR method’s fault detection and false alarm rates under different relative error thresholds. The results are shown in the left plot of Figure~\ref{fig:SNVR_coverage}. By analyzing the numerical difference between the checksum and the actual result in both error-free and error-injected scenarios, we found that when the relative error threshold is set to 7e-6, the fault detection rate is around 97.2\% and the false alarm rate is 5.9\%, maximizing the error coverage. Additionally, SNVR protects the numerator and denominator separately in softmax computation, while the traditional restriction method only safeguards the final division result. This gives SNVR a higher error coverage rate and a lower correction error. The error distribution after applying restriction methods is plotted on the right side of Figure~\ref{fig:SNVR_coverage}. The results show that the SNVR method restricts most errors within the range of 0 to 0.02. In contrast, traditional methods cannot confine errors within a small range, with values widely distributed between 0 and 0.15, leading to greater uncertainty in fault tolerance.

\subsection{Unified Verification and Fault Tolerance in Transformers}
In this subsection, we first evaluate the total execution time and fault tolerance overhead of the optimized end-to-end fault tolerant attention with unified verification. We then simulate the overhead of error detection and correction using the optimized EFTA on common Transformer models.

\textbf{\large \textbullet{ }}\textbf{EFTA with Unified Verification.} We further optimized the EFTA using a unified verification strategy, encoding multi-step computations into a single tensor checksum for error detection. As shown in Table~\ref{tab:opt-1}, when 16 attention heads are used and the head dimension is set to 64, the optimized EFTA reduces the average fault tolerance overhead to 15.3\% compared to 53\% in the unoptimized version, achieving a 1.32$\times$ speedup on average. Furthermore, compared to the traditional decoupled method, optimized EFTA achieves an average speedup of 7.56$\times$.

For the attention setting of large models with head = 32 and dim = 64, the optimized EFTA incurs an average overhead of 12.5\% for fault tolerance, compared to 22.7\% in the unoptimized version. Our method is on average 3.69$\times$ faster than existing decoupled fault tolerance approaches. The corresponding results are presented in Table~\ref{tab:opt-2}.

\small
\begin{table}[t]
    \vspace{-5pt}
    \captionsetup{justification=centering}
    \caption{Comparison of EFTA and optimized EFTA for \\ head = 16, dim = 64}
    \vspace{-8pt}
    \centering
    \begin{tabular}{lcccc}
    \toprule
    \textbf{Length} & \textbf{EFTA} (ms) & \textbf{Overhead} & \textbf{EFTA-o} (ms)& \textbf{ Overhead} \\
    \midrule
    512    & 0.425 & 52.3\% & 0.315 & 12.9\% \\
    1k & 0.924  & 40.2\% & 0.718 & 8.9\% \\
    2k   & 1.537 & 48.0\% & 1.178 & 13.4\% \\
    4k    & 2.924 & 66.5\% & 2.004 & 14.1\% \\
    8k    & 4.966 & 62.9\% & 3.951 & 29.6\% \\
    16k    & 13.804 & 48.2\% & 10.507 & 12.8\% \\
    \bottomrule
    \end{tabular}
    \label{tab:opt-1}
    \vspace{-3pt}
\end{table}
\normalsize

\small
\begin{table}[t]
    \vspace{-3pt}
    \captionsetup{justification=centering}
    \caption{Comparison of EFTA and optimized EFTA for \\ head = 32, dim = 128}
    \vspace{-8pt}
    \centering
    \begin{tabular}{lcccc}
    \toprule
    \textbf{Length} & \textbf{EFTA} (ms) & \textbf{Overhead} & \textbf{EFTA-o} (ms)& \textbf{ Overhead}  \\
    \midrule
    512    & 1.498 & 24.9\% & 1.199 & 13.4\%  \\
    1k & 2.810  & 24.7\% & 2.253 & 13.5\% \\
    2k   & 5.441 & 24.6\% & 4.364 & 13.4\% \\
    4k    & 10.703 & 26.1\% & 8.483 & 14.8\% \\
    8k    & 18.912 & 27.0\% & 14.886 & 15.4\% \\
    16k    & 32.728 & 9.1\% & 29.995 & 4.5\% \\
    \bottomrule
    \end{tabular}
    \label{tab:opt-2}
\end{table}
\normalsize

\begin{figure}[h]
\centerline{\includegraphics[scale=0.58]{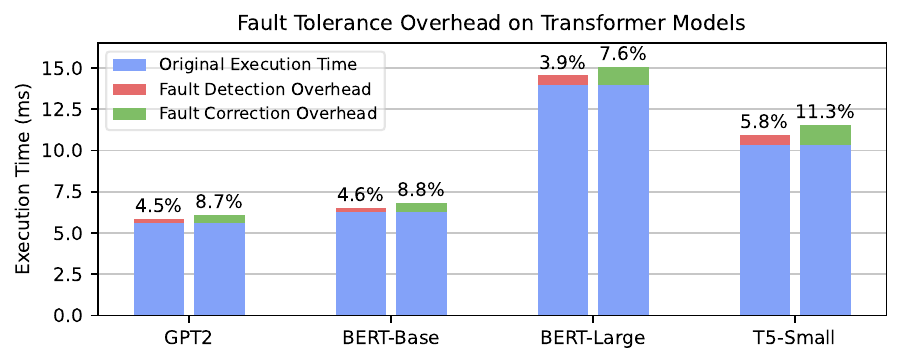}}
\vspace{-5pt}
\caption{Error detection and correction overhead of EFTA on common Transformer models.}
\vspace{-5pt}
\label{fig:model_simu}
\Description{model_simu}
\end{figure}

\textbf{\large \textbullet{ }}\textbf{EFTA on Transformer Models.} To further investigate the fault tolerance performance of our end-to-end fault-tolerant attention in practical applications, we simulate the error detection and correction overhead of applying optimized EFTA on common transformer models, including GPT2, BERT-Base, BERT-Large, and T5-Small. During the experiments, we fixed the input length at 512 and set the hidden dimension and layer counts of EFTA according to the model configuration. We profile the execution time for each inference step and the simulation results show that the GPT2 model takes approximately 5.6 ms to generate one token. This time increases to 5.874 ms when EFTA is used. Figure~\ref{fig:model_simu} provides a detailed breakdown of execution time and overhead. For error detection, EFTA incurs an average overhead of 4.7\% across the four models. For error correction, we simulate errors by injecting a single bit flip for each attention computation, and correcting these errors incurs an average overhead of 9.1\%.

\section{Conclusion}
In this paper, we propose an end-to-end fault tolerant attention (EFTA) to enhance the reliability of Transformer model inference against soft errors. The proposed EFTA fuses attention and fault tolerance computations within a single kernel, mitigating redundant memory access. In addition, an architecture-aware and computation-optimized hybrid fault tolerance scheme is incorporated in EFTA, enabling efficient error detection and correction for mixed-precision matrix operations and softmax computations on Tensor Cores. Experimental results show that our EFTA achieves up to 7.56$\times$ speedup compared to existing decoupled methods with a 13.9\% average fault tolerance overhead. Further simulations on Transformer models show that EFTA adds negligible overhead to the inference process - 4.7\% for error detection and 9.1\% for error correction.



\bibliographystyle{ACM-Reference-Format}
\bibliography{sample-base}

\appendix








\end{document}